\documentclass[cits]{PoS}

\title{Analytic continuation in QCD-like theories at finite density and finite isospin}

\ShortTitle{Analytic continuation in QCD-like theories}

\author{P. Cea\\
            Universit\`a di Bari \& INFN - Bari, Italy\\
            E-mail: \email{paolo.cea@ba.infn.it}}
           
\author{\speaker{L. Cosmai}\\
        INFN - Bari, Italy\\
        E-mail: \email{leonardo.cosmai@ba.infn.it}}
            
\author{M. D'Elia\\
            Universit\`a di Genova \& INFN - Genova, Italy\\
            E-mail: \email{massimo.delia@ge.infn.it}}

\author{A. Papa\\
            Universit\`a della Calabria \& INFN - Cosenza, Italy\\
            E-mail: \email{papa@cs.infn.it}}

\abstract{The method of analytic continuation is one of the most powerful tools
to circumvent the sign problem in lattice QCD. The present study is part of a larger project which, based on the
investigation of QCD-like theories which are free of the sign problem, is
aimed at testing the validity of the method of analytic
continuation and at improving its predictivity, in view of its
application to real QCD.
We have shown that a
considerable improvement can be achieved if suitable functions are
used to interpolate data with imaginary chemical potential. We present
results obtained in a theory free of the sign problem such as two-color
QCD at finite chemical potential.
}

\FullConference{The XXVII International Symposium on Lattice Field Theory - LAT2009\\
		 July 26-31 2009\\
		 Peking University, Beijing, China}

\begin{document}

\section{Introduction}
The investigation the phase diagram of QCD in the temperature-chemical potential 
plane has a deep relevance and implications on cosmology, astrophysics and in the phenomenology of heavy
ions collisions. 
The lattice formulation of QCD is the only tool to approach this important issue starting
from first principles.  Unfortunately the study of QCD at non-zero baryonic density by numerical simulations on a
space-time lattice is plagued by the well-known sign problem: the fermion
determinant is complex and the Monte Carlo sampling becomes unfeasible.
One of the possibilities to circumvent this problem is to perform Monte
Carlo numerical simulations for imaginary values of the baryonic chemical
potential, where the fermion determinant is real and the sign problem is
absent, and to infer the behavior at real chemical potential by analytic
continuation.
The method of analytic continuation~\cite{deForcrand:2002ci,deForcrand:2003hx,D'Elia:2002gd,D'Elia:2004at,Azcoiti:2005tv,Chen:2004tb,Giudice:2004se,Kim:2005ck,Cea:2006yd,Papa:2006jv,deForcrand:2006pv,deForcrand:2007rq,D'Elia:2007ke,Conradi:2007be,Cea:2007wa,Karbstein:2006er} is well-founded and works fine within
the limitations posed by the presence of non-analyticities and by the periodicity of
the theory with imaginary chemical potential~\cite{Roberge:1986mm}.

It is very important to answer the question about which is the optimal way to extract 
information from data taken at imaginary values of the chemical potential.
This is equivalent to answer which is the best interpolating function for data
at imaginary chemical potential that could analytically continued in order to 
get physical predictions for real values of $\mu$. 
The aim of our investigations in the past two years has been to to study limitations and
possible improvements of the method of analytic continuation~\cite{Cea:2006yd,Cea:2007vt,Cea:2009ba}.
In order to study this problem we have considered SU(2) (two-color QCD)  and SU(3) at finite isospin. 
Indeed these theories
are free of the sign problem and Monte Carlo numerical simulations at real values of the chemical potential 
or at real values of isospin potential  are feasible. Therefore it is possible to compare the analytic continuations 
with the data from direct simulations allowing at the same time to discriminate between interpolating functions an to 
test the range of the reliability of the analytic continuation.
Here we briefly review results obtained in studying analytic continuation of physical observables and of the critical line 
in two-color QCD. Results obtained for SU(3) at finite isospin has been reviewed in ref.~\cite{AlexLat2009}.

\section{Analytic continuation of physical observables}

As already shown long ago~\cite{Roberge:1986mm},
the partition function of any SU(N) gauge theory with non-zero
temperature and imaginary chemical potential, $\mu=i\mu_I$, is
periodic in $\theta\equiv\mu_I/T$ with period $2\pi/N$ and that the
free energy $F$ is a regular function of $\theta$ for $T < T_E$,
while it is discontinuous at $\theta=2\pi(k+1/2)/N$,
$k=0,1,2,\ldots$, for $T > T_E$, where $T_E$ is a characteristic
temperature, depending on the theory. 

We have considered SU(2) in presence of $n_f=8$ degenerate staggered fermions of mass $am=0.07$.
Figure 1 shows the the tentative phase diagram in the $(\mu_I,\beta)$ plane for this theory (in correspondence of a fermion mass $am=0.07$) 
with $\beta_E \simeq 1.55$~\cite{Gottlieb:1988cq,Giudice:2004pe}
and $\beta_c \simeq 1.41$~\cite{Liu:2000in}.
We have performed numerical simulations on a $16^3 \times 4$ lattice using the exact $\phi$ algorithm~\cite{Gottlieb:1988cq}  
with $dt=0.01$ (typical statistics: ~ 20k trajectories). 
We have performed a careful test of  the analytic continuation of physical observables. 
A detailed discussion of the results obtained is reported in ref.~\cite{Cea:2006yd}. 
In order to show the importance of  a careful choice of the interpoIating function 
of the  imaginary chemical potential data,  in figure~\ref{fig2} we display results for two different observables in correspondence
of two different values of $\beta$.  We have shown that the use of ratio of polynomials as  interpolating function 
can lead to a dramatic improvement in the analytic continuation of physical observables.

The aim of our subsequent investigations has been to understand if a  careful choice of the interpolating function can also improve
the continuation of the critical line.

\begin{figure}
\centering
\includegraphics[width=.5\textwidth,clip]{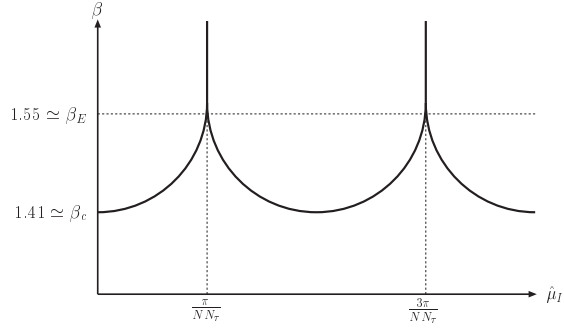}
\caption{Phase diagram in the $(\beta,\hat\mu_I)$-plane; $N$ is
the number of colors, $N_\tau$ the extension of the lattice in the
temporal direction. The numerical values for $\beta_E$ and $\beta_c$
are valid for SU(2) in presence of $n_f=8$ degenerate staggered
fermions with mass $am=0.07$}
\label{fig1}
\end{figure}

\begin{figure}[htbp]
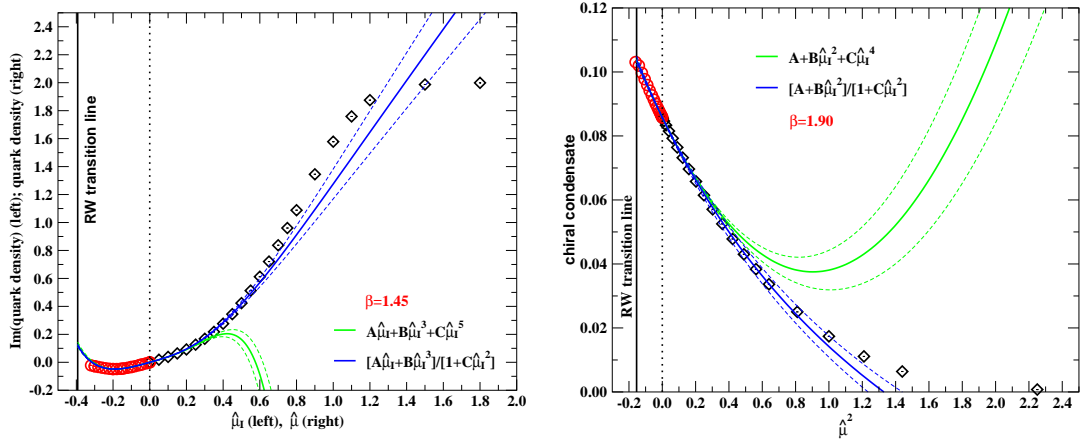
 
\centering 
\begin{minipage}[c]{.48\textwidth}
\centering
\includegraphics[width=.95\textwidth,clip]{final_b145_ndens_new.eps} 
\end{minipage}%
\hspace{1mm}%
\begin{minipage}[c]{.48\textwidth}
\centering
\includegraphics[width=.95\textwidth,clip]{final_b190_psibpsi_new.eps} 
\end{minipage} 
\caption{(Left) Negative side of the horizontal axis: imaginary
part of the fermion number density {\it vs.} the imaginary
chemical potential at $\beta=1.45$. Positive side of the horizontal
axis: real part of the fermion number density {\it vs.} the real
chemical potential at $\beta=1.45$.
The green (blue) solid lines represent the polynomial (ratio of polynomials)
interpolating function; the dashed lines give the corresponding uncertainty,
coming from the errors in the parameters of the fit.
(Right) Chiral condensate {\it vs.} $\mu^2$ at $\beta=1.90$.
The green (blue) solid lines represent the polynomial (ratio of polynomials)
interpolating function; the dashed lines give the corresponding uncertainty,
coming from the errors in the parameters of the fit.
\label{fig2}} 
\end{figure}



\section{Analytic continuation of the critical line}

The determination of the critical line in the $(T,\mu)$ plane is of overwhelming importance 
for the study of strong interactions at finite temperature and baryon density.
The analytic continuation of the (pseudo-)critical line on the temperature-chemical potential plane 
is well-justified, but a careful test in two-color QCD
and three-color QCD with finite isospin chemical potential has cast some
doubts on its reliability~\cite{Cea:2007vt,Cea:2009ba,AlexLat2009}.

\begin{figure}
\centering
\includegraphics[width=.4\textwidth,clip]{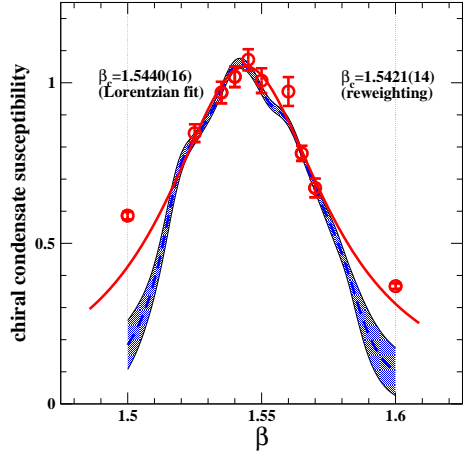}
\caption{Chiral susceptibility at $(a \mu)^2=-0.1225$ vs. $\beta$. Full red line is the Lorentzian
fit. Dashed blue line is the multihistogram reweighting within its bootstrap error (blue strip). }
\label{fig3}
\end{figure}

\TABLE{
\setlength{\tabcolsep}{0.5pc}
\centering
\caption[]{Summary of the values of $\beta_c(\mu^2)$ obtained by fitting the 
peaks of the susceptibilities of chiral condensate 
$\langle\overline\psi\psi\rangle$, Polyakov loop $\langle L \rangle$ and 
plaquette $\langle P \rangle$ in SU(2) on a 16$^3\times 4$ lattice with
fermion mass $am$=0.07. For each interpolation the $\chi^2/{\rm d.o.f.}$
is given.}
\begin{tabular}{rrclclc}
\hline
\hline
\multicolumn{1}{c}{\hspace{0.70cm}$(a\mu)^2$} &
\multicolumn{1}{c}{\hspace{1cm}$\langle\overline\psi\psi\rangle$} &
$\chi^2/{\rm d.o.f.}$ &
\multicolumn{1}{c}{\hspace{1cm}$\langle L \rangle$} &
$\chi^2/{\rm d.o.f.}$ &
\multicolumn{1}{c}{\hspace{1cm}$\langle P \rangle$} &
$\chi^2/{\rm d.o.f.}$ \\
\hline
-0.1225 & 1.5440(16) & 1.34 & 1.5349(43) & 0.85 & 1.5418(24) & 0.93 \\
-0.09   & 1.5068(15) & 0.65 & 1.5019(29) & 0.25 & 1.5046(21) & 1.06 \\
-0.0625 & 1.4775(29) & 0.88 & 1.4665(32) & 0.31 & 1.4755(37) & 0.65 \\
-0.04   & 1.4532(16) & 0.50 & 1.4453(36) & 0.76 & 1.4522(26) & 1.21 \\
-0.0225 & 1.4324(22) & 1.20 & 1.4240(28) & 0.80 & 1.4300(39) & 0.80 \\
-0.01   & 1.4197(16) & 1.86 & 1.4104(33) & 0.43 & 1.4199(26) & 1.45 \\
 0.     & 1.4102(18) & 0.07 & 1.3989(61) & 0.49 & 1.4117(32) & 0.07 \\
 0.04   & 1.3528(22) & 2.91 & 1.3388(72) & 1.01 & 1.3631(46) & 1.16 \\
 0.0625 & 1.3145(30) & 1.34 & 1.2976(62) & 0.87 & 1.3286(50) & 1.28 \\
 0.09   & 1.2433(59) & 1.09 & 1.2508(62) & 0.98 & 1.2864(109)& 0.60 \\
\hline
\hline
\end{tabular}
\label{table1}
}

In this section we present  our results in the determination of the critical line  in two-color QCD
using the method of analytic continuation.
Contrary to the case of physical observables discussed in the previous section, the theoretical 
basis is not straightforward since it relies on the assumption that susceptibilities, whose peak signals the presence of the transition,
be analytic functions of the parameters on a finite volume~\cite{deForcrand:2002ci,deForcrand:2003hx}.
We have tested the method of analytic continuation in the case of two-color QCD and in the case of 
QCD ad finite isospin density~\cite{Cea:2007vt,Cea:2009ba}. 
As for usual QCD simulations, we have determined the critical line for imaginary values of the chemical potential ($\mu^2 < 0$) and
interpolate them by suitable functions to be continued to $\mu^2 > 0$. In order to test the reliability of the analytic continuation, the prediction obtained at real $\mu$ has been compared with direct determinations of the transition line.

\begin{figure}[htbp]
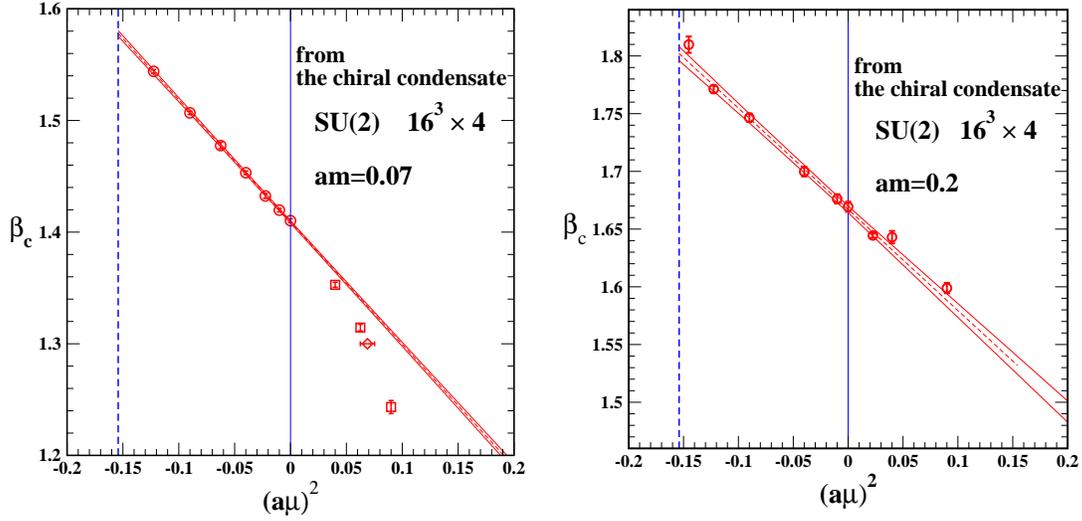
 
\centering 
\begin{minipage}[c]{.48\textwidth}
\centering
\includegraphics[width=.95\textwidth,clip]{psibpsi_007.eps} 
\end{minipage}%
\hspace{1mm}%
\begin{minipage}[c]{.48\textwidth}
\centering
\includegraphics[width=.95\textwidth,clip]{psibpsi_02.eps} 
\end{minipage} 
\caption{Critical couplings obtained from the susceptibility of
chiral condensate in SU(2) on a 
16$^3\times 4$ lattice with $am$=0.07 (left) and $am$=0.2 (right), together
with a linear fit (dotted line) in $(a\mu)^2$ to data with $\mu^2 \leq 0$.
The solid lines around the 
best fit line delimit the 95\% CL region.
\label{fig4}} 
\end{figure}

\begin{figure}[htbp]
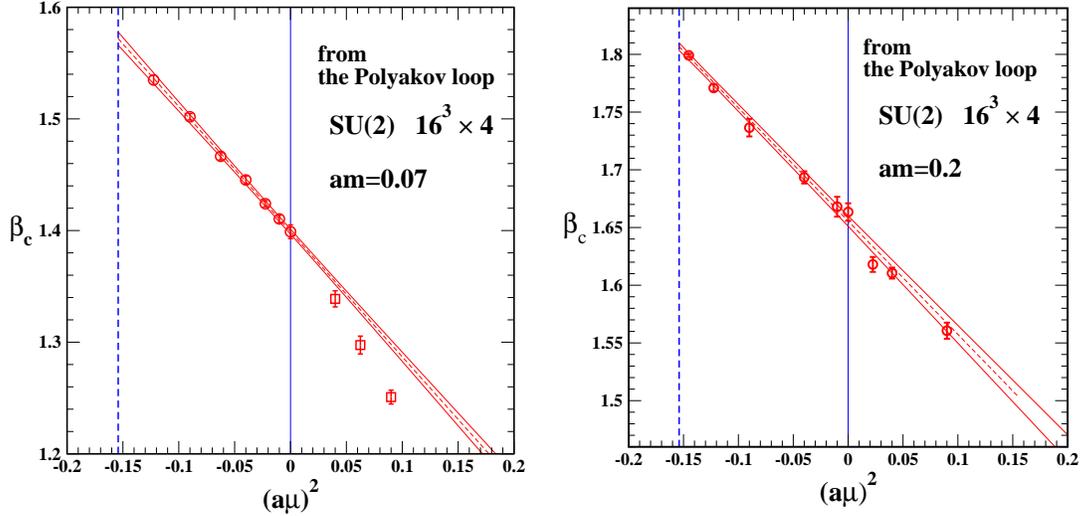

\centering 
\begin{minipage}[c]{.48\textwidth}
\centering
\includegraphics[width=.95\textwidth,clip]{poly_007.eps} 
\end{minipage}%
\hspace{1mm}%
\begin{minipage}[c]{.48\textwidth}
\centering
\includegraphics[width=.95\textwidth,clip]{poly_02.eps} 
\end{minipage} 
\caption{Critical couplings obtained from the susceptibility of
Polyakov loop  in SU(2) on a 
16$^3\times 4$ lattice with $am$=0.07 (left) and $am$=0.2 (right), together
with a linear fit (dotted line) in $(a\mu)^2$ to data with $\mu^2 \leq 0$.
The solid lines around the 
best fit line delimit the 95\% CL region.
\label{fig5}} 
\end{figure}

It should be remarked that on a finite volume there are no true nonanalyticities, therefore the location of the critical line may depend on 
the observable chosen to probe the transition. Consequently we have determined the (pseudo-)critical coupling $\beta_c(\mu^2)$ by looking at 
the peaks of the susceptibilities of three different observables: the chiral condensate, the Polyakov loop, and the plaquette.
In figure~\ref{fig3} we show an example of this determination.  We have fitted the peak according to a Lorentzian function. The result obtained in this way  for $\beta_c(\mu^2)$ agrees well with the result obtained by means of the multihistogram reweighting.

In table~\ref{table1} the values of $\beta_c(\mu^2)$ obtained in correspondence of the different 
"probe"  observable (at fermion mass $am=0.07$)  are shown. They depend very weakly on the observable considered 
(only in one case the relative deviation between two determinations at the
same $\mu^2$ slightly exceed $3 \sigma$). 
The strategy is now to interpolate the critical $\beta$'s obtained
at fixed imaginary chemical potential with
an analytic function of $\mu$, to be then
extrapolated to real chemical
potential.  For a theory free from the sign
problem (such as two-color QCD) the extrapolated
curve can be compared with the determinations of the
critical $\beta$'s at real chemical potential obtained through a direct computation.

\begin{figure}
\centering
\includegraphics[width=.5\textwidth,clip]{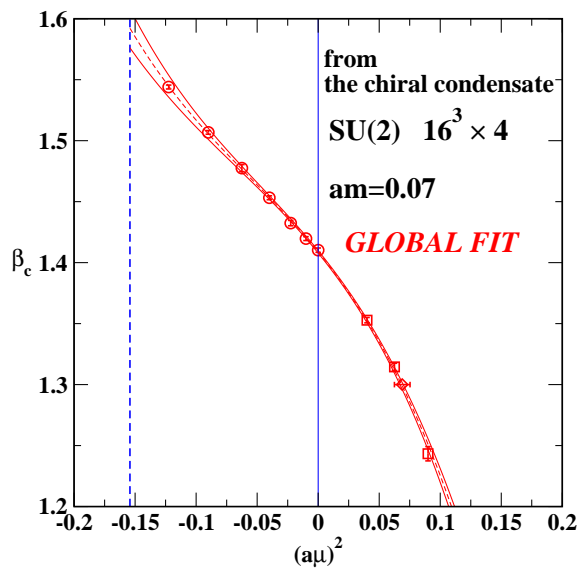}
\caption{Same as in figure~4, but with results of a fit
to all data including term up to order $\mu^6$.
\label{fig6}}
\end{figure}

As displayed on the left sides  of figure~\ref{fig4} and figure~\ref{fig5},
imaginary chemical potential data are very well fitted with a linear polynomial 
in $\mu^2$. 
If different functions are used (larger order polynomials,
ratio of polynomials) the fit puts to value compatible with
zero all parameters (except two of them) thus reducing
again to a first order polynomial in $\mu^2$. 
This is in marked difference with what we found (see previous section and ref.~\cite{Cea:2006yd})
for the analytic continuation of physical observables, where the ratio of polynomials performed very well.
Moreover, on the left sides of figures~\ref{fig4} and~\ref{fig5},  we can clearly see a quite significative
deviation between extrapolation and direct determination of the critical line at real chemical potential.
The discrepancy found could imply that or the the critical line is not analytic on the whole interval of $\mu^2$
or that the interpolation at $\mu^2 \le 0$ is not accurate enough to correctly reproduce the behavior at $\mu^2 > 0$.
Indeed we have found (see figure~\ref{fig6}) that a polynomial of third order in $\mu^2$ nicely fits all data for $\beta_c(\mu^2)$
and therefore the critical line is analytic on the whole interval of $\mu^2$.
A possible conclusion is that 
for $\mu^2 \le 0$ the $\mu^4$ and $\mu^6$ terms compensate each other at large negative values
of  $\mu^2$ so that the effective interpolating function of the data at $\mu^2 \le 0$  is a first
order polynomial in $\mu^2$  while 
for $\mu^2 \ge 0$  the $\mu^4$ and $\mu^6$ terms work in the same direction and their contribution
cannot be neglected.

In order to  verify  if this scenario is peculiar to SU(2) we have investigated the same theory 
with a different mass $am=0.2$ for  the $n_f=8$ degenerate fermions.
We have also examined the case of SU(3) at finite isospin~\cite{Cea:2009ba,AlexLat2009}.
The reason for consider SU(2) at  a larger value for the quark mass is that at any fixed temperature $T$
the critical value of $\mu$ gets larger, consequently the critical line could become less curved in
the physical region $\mu^2 > 0$. Accordingly higher order terms in $\mu^2$ in the description of the critical line
by a polynomial could be less important.
We have sampled the critical line for SU(2) and quark mass $am=0.2$. 
The best interpolation of   $\beta_c(\mu^2)$ data at $\mu^2 \le0$  is a polynomial  linear in $\mu^2$ 
and, at variance with the case of quark mass $am=0.07$, 
the extrapolation to $\mu^2 >0$ compares very well with the direct determination of $\beta_c(\mu^2)$ in that region.
So we can argue that the the extrapolation to $\mu^2>0$ works definitely better for larger quark masses, i.e. away form the
chiral limit.  

\section{Conclusions}

We have reported here an investigation which is part of a larger project devoted to study the method of analytic
continuation in QCD-like theories free from the sign problem and to improve its predictivity in view of its application to QCD.

For what concerns analytic continuation of physical observables we have shown that a considerable improvement can be 
achieved, when extrapolating data from imaginary to real chemical potentials, if ratios of polynomials are used as interpolating functions
(for a thorough discussion see ref.~\cite{Cea:2006yd}).

We have also presented results for the analytic continuation of the critical line in the $(T,\mu)$  plane from imaginary to real
chemical potential both in the case of two-color QCD.
We have found that the critical line around $\mu=0$ can be described by an analytic function.
Indeed, a third order polynomial in $\mu^2$ fits all the available data for the critical coupling.

We have shown that  there is a clear indication that in the chiral limit high-order terms in the polynomial
interpolation play a relevant role at $\mu^2 >0$  but are less visible at $\mu^2<0$, this calling for extremely 
high accuracy in detecting such terms from simulations at $\mu^2 < 0$. 
The predictions for the pseudocritical couplings at real chemical potentials may be wrong if data at imaginary $\mu$ are 
fitted according to a linear dependence. 

All the issues  above have undergone further investigation in a different theory such as SU(3) at finite isospin
(results for SU(3) at finite isospin are discussed in refs.~\cite{Cea:2009ba,AlexLat2009}). 

The lessons we learned  in studying analytic continuation in QCD-like theory free from the sign problem will be used in the near future
to the determination the critical line for real QCD.




\providecommand{\href}[2]{#2}\begingroup\raggedright\endgroup

\end{document}